\documentclass[twocolumn,amsmath,amssymb,aps,preprintnumbers,showpacs,superscriptaddress,floatfix]{revtex4}

\usepackage{bm}         
\usepackage{amssymb}
\usepackage{amsmath}
\usepackage{graphicx}
\usepackage{epsfig}
\usepackage{color}
\usepackage{ulem}
\usepackage{SIunits}
\usepackage{epstopdf}
\usepackage[USenglish]{babel}

\begin{document}

\title{Observing electron spin resonance between 0.1 and 67~GHz at temperatures between 50~mK and 300~K using broadband metallic coplanar waveguides}

\author{Yvonne~Wiemann}
\affiliation{1.~Physikalisches Institut, Universit\"{a}t Stuttgart, Pfaffenwaldring 57, 70569 Stuttgart, Germany}
\author{Julian~Simmendinger}
\affiliation{1.~Physikalisches Institut, Universit\"{a}t Stuttgart, Pfaffenwaldring 57, 70569 Stuttgart, Germany}
\author{Conrad~Clauss}
\affiliation{1.~Physikalisches Institut, Universit\"{a}t Stuttgart, Pfaffenwaldring 57, 70569 Stuttgart, Germany}
\author{Lapo~Bogani}
\affiliation{1.~Physikalisches Institut, Universit\"{a}t Stuttgart, Pfaffenwaldring 57, 70569 Stuttgart, Germany}
\affiliation{Department of Materials, University of Oxford, 16 Parks Road, Oxford OX1 3PH, United Kingdom}
\author{Daniel~Bothner}
\affiliation{Physikalisches Institut and Center for Collective Quantum Phenomena in LISA$^+$, Universit\"{a}t T\"{u}bingen, Auf der Morgenstelle 14, 72076 T\"{u}bingen, Germany}
\author{Dieter~Koelle}
\affiliation{Physikalisches Institut and Center for Collective Quantum Phenomena in LISA$^+$, Universit\"{a}t T\"{u}bingen, Auf der Morgenstelle 14, 72076 T\"{u}bingen, Germany}
\author{Reinhold~Kleiner}
\affiliation{Physikalisches Institut and Center for Collective Quantum Phenomena in LISA$^+$, Universit\"{a}t T\"{u}bingen, Auf der Morgenstelle 14, 72076 T\"{u}bingen, Germany}
\author{Martin~Dressel}
\affiliation{1.~Physikalisches Institut, Universit\"{a}t Stuttgart, Pfaffenwaldring 57, 70569 Stuttgart, Germany}
\author{Marc~Scheffler}
\affiliation{1.~Physikalisches Institut, Universit\"{a}t Stuttgart, Pfaffenwaldring 57, 70569 Stuttgart, Germany}

\date{\today}

\begin{abstract}
We describe a fully broadband approach for electron spin resonance (ESR) experiments where it is possible to not only tune the magnetic field but also the frequency continuously over wide ranges. Here a metallic coplanar transmission line acts as compact and versatile microwave probe that can easily be implemented in different cryogenic setups. We perform ESR measurements at frequencies between 0.1 and 67~GHz and at temperatures between 50~mK and room temperature. Three different types of samples (Cr$^{3+}$ ions in ruby, organic radicals of the nitronyl-nitroxide family, and the doped semiconductor Si:P) represent different possible fields of application for the technique.
We demonstrate that an extremely large phase space in temperature, magnetic field, and frequency for ESR measurements, substantially exceeding the range of conventional ESR setups, is accessible with metallic coplanar lines.
\end{abstract}

\pacs{87.80.Lg, 76.30.Rn, 84.40.Az, 07.57.Pt}

\maketitle

ESR spectroscopy is used to investigate magnetic material properties in many fields of natural sciences. Typically, a static magnetic field $B$ is swept while a fixed-frequency microwave or radio frequency (RF) magnetic field irradiates the sample material. If $B$ and microwave frequency $\nu$ fulfill the resonance condition
\begin{eqnarray}
h\nu=g\mu_{B}B
\end{eqnarray}
with $h$ Planck's constant and $\mu_{B}$ the Bohr magneton, resonant power absorption can be observed due to magnetic transitions between different energy states that are governed by the external magnetic field (Zeeman splitting). Typical commercial ESR setups operate in a very narrow frequency range or even at a single frequency, e.g.\ around 10~GHz or 34~GHz for X- and Q-band spectrometers, respectively \cite{Poole97}.
While such instrumentation proved to be extremely useful in a wide range of conventional ESR studies in physics, biology, and chemistry, it can show limitations in the study of less conventional magnetic materials for example those with broad lineshapes, large zero-field splitting, or numerous magnetic transitions. Here, measurements at a single ESR frequency might not be sufficient to elucidate the level structure.
An additional complication can arise if field-induced spin level (anti-)crossings occur in systems with low-lying spin excited states \cite{Slageren2002, Choi06}. Here, a low ESR frequency is combined with a large static magnetic field beyond the range of conventional ESR electromagnets.
The growing interest in ESR experiments beyond commercial single-frequency setups is evident in particular for higher frequencies where multifrequency or broadband ESR instrumentation in a wide range of magnetic field has already been implemented \cite{vanSlageren03, Smith1998, Hassan2000}. Several broadband magnetic resonance studies at lower GHz frequencies have been limited to selected materials that could easily be implemented \cite{Schwartz2000,Giesen2005,Schuster2010} while others reached a certain frequency tunability by changing the dimension of three-dimensional ESR cavities \cite{Narkowicz05,Narkowicz08,Malissa13,Reijerse10, Gatteschi02, vanSlageren03}. But also other homemade ESR spectrometers with features such as broadband operation have been constructed and reported \cite{Denysenkov03,Harward11}. While all these setups might be very helpful for many investigated materials, they all have limitations when it comes to the actually demonstrated frequency range, application at large magnetic fields, or cryogenic temperatures.

\begin{figure}[tb]		
	\centering
	\includegraphics[width=8cm]{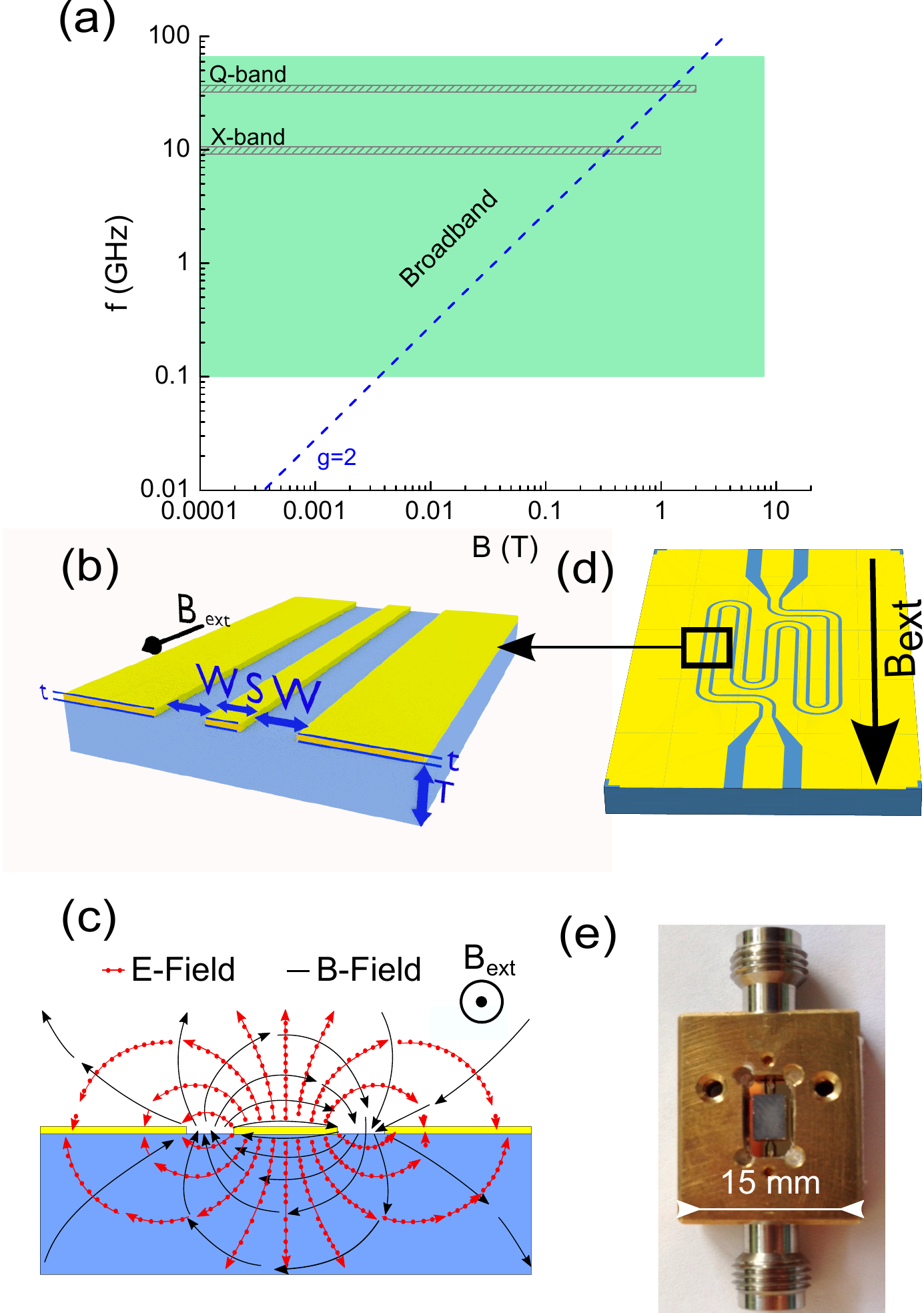}
	\caption{(color online) (a) Schematic illustration of the frequency and magnetic field range that can be reached with the broadband ESR technique (green area) presented in this paper. The hatched grey boxes represent the area that can be reached with typcial commercial X- and Q-band ESR spectrometers whereas the blue dashed line indicates a spin system with $g=2$. (b) Schematic view of the metallic waveguide with characteristic dimensions ($T$: sapphire thickness $=430\,\micro\meter$, $t$: overall thickness of metallic films $=305\,\nano\meter$, $W$: distance between center conductor and ground planes, $S$: center conductor width). (c) Orientation of the magnetic (black line) and electric (dotted red line) RF fields between center conductor and ground planes. (d) Layout of the waveguide chip used for the experiment. (e) Photo of the sample box with mounted waveguide and sample (here Si:P).}
	\label{fig:Figure1}
\end{figure}

In this paper, we demonstrate broadband ESR measurements that cover the extremely broad frequency range from $0.1$ up to $67\,\giga\hertz$ in single frequency sweeps and that span temperatures from 300~K down to $50\,\milli\kelvin$. 
Fig.\ \ref{fig:Figure1}(a) compares the frequency and magnetic field ranges that are accessed by conventional ESR spectrometers (grey hatched boxes) such as X- and Q-bands as the most common ones, and the broadband ESR technique (green box) presented in this work. The blue dashed line indicates a spin system with $g=2$. Clearly, our approach covers a much larger phase space than conventional ESR setups. Furthermore, our ESR probe can conveniently be operated in commercial superconducting magnets and thus could directly be applied at even larger fields than the 8~T of our magnet, if particular scientific question might require this. This is similar to conventional ESR spectrometers that have been equipped with superconducting magnets in special cases. But more important, while the sample temperature of commercial ESR setups as well as of most of the dedicated special approaches mentioned above cannot be lower than a few K, our probe can conveniently be installed in commercial dilution refrigerators, which enables sample temperature for ESR studies well below 100 mK.

In our approach, microfabricated metallic (Au and Cu) coplanar waveguides are used to confine the RF magnetic field. In a previous study \cite{Clauss13}, we utilized superconducting coplanar waveguides up to 40~GHz. Superconducting transmission lines in principle exhibit lower intrinsic microwave losses at cryogenic temperatures than metallic lines, but are also prone to disadvantages. In particular, superconducting lines become quite lossy when operated above the critical temperature, and they show strong temperature and magnetic field dependence of the losses, which seriously influences the data analysis. Moreover, the static magnetic field can be deformed substantially by the superconducting structure resulting in frequency-shifted ESR absorption \cite{Clauss13}.

Here, we show how these problems are overcome by using metallic waveguides. We demonstrate the broad applicability of our method using three diverse exemplary materials with established ESR properties, namely a ruby single crystal [Al$_2$O$_3$:Cr$^{3+}$], the doped semiconductor Si:P, and a nitronyl-nitroxide radical  2-(4'-p-methoxyphenyl)-4,4,5,5-tetramethylimidazoline-1-oxyl-3-oxide (NITPhOMe for brevity) \cite{Benelli1990, Caneschi01}, which we investigate in a wide range of temperatures and ESR frequencies.

Fig.\ \ref{fig:Figure1}(b) shows the main structure of a metallic coplanar waveguide. The central conductor (of width $S$) carries the RF signal and is centered, at a distance $W$, between two ground planes, which act as outer conductor. Unless stated otherwise, we use a coplanar waveguide of $S=60\,\micro\meter$ and $W=25\,\micro\meter$. The whole chip consists of three different layers. On top of a  $430\,\micro\meter$ thick r-cut sapphire substrate, a $5\,\nano\meter$ thick chromium adhesion layer is evaporated and followed by $300\,\nano\meter$ of sputtered Au or Cu. The films are then structured by UV lithography. The final chip has a size of $7\times 4\,\milli\meter^2$ (see Fig. \ref{fig:Figure1}(d)) and is placed into a specifically provided gold-plated brass box. This compact design is compatible with implementation in a superconducting magnet as well as in a dilution refrigerator for mK studies \cite{Scheffler13}, and one can easily perform broadband ESR measurements in different setups with exactly the same probe.
Silver paste establishes good electrical contact to the 1.85~mm-type coaxial connectors (center conductor) and to the box (ground planes). The different samples were directly placed onto the chip. In case of ruby and Si:P, the cuboid-shaped sample was fixed with vacuum grease (see Fig.\ \ref{fig:Figure1}(e)) whereas the organic radical NITPhOMe was dissolved in isopropyl and then drop-by-drop transferred to the waveguide. Evaporation of the solvent leads to crystallization directly on the waveguide.

Typically, the sample box is attached to a sample holder which is embedded into a magnet cryostat and can be cooled down to $1.6\,\kelvin$. As shown in Fig.\ \ref{fig:Figure1}(b), the static magnetic field is oriented parallel to the film \cite{filmorientation}. Spin flips causing ESR can only be induced if the magnetic field component of the microwave signal (see Fig. \ref{fig:Figure1}(c)) and the static external magnetic field are perpendicular to each other (see Fig.\ \ref{fig:Figure1}(c)).
A continuous microwave signal is sent via coaxial cables to the coplanar waveguide in the magnet, and the transmitted power is detected by a power meter or a network analyzer. The input power was adapted to each measurement \cite{power}.

\begin{figure}[tb]		
	\centering
	\includegraphics[width=8cm]{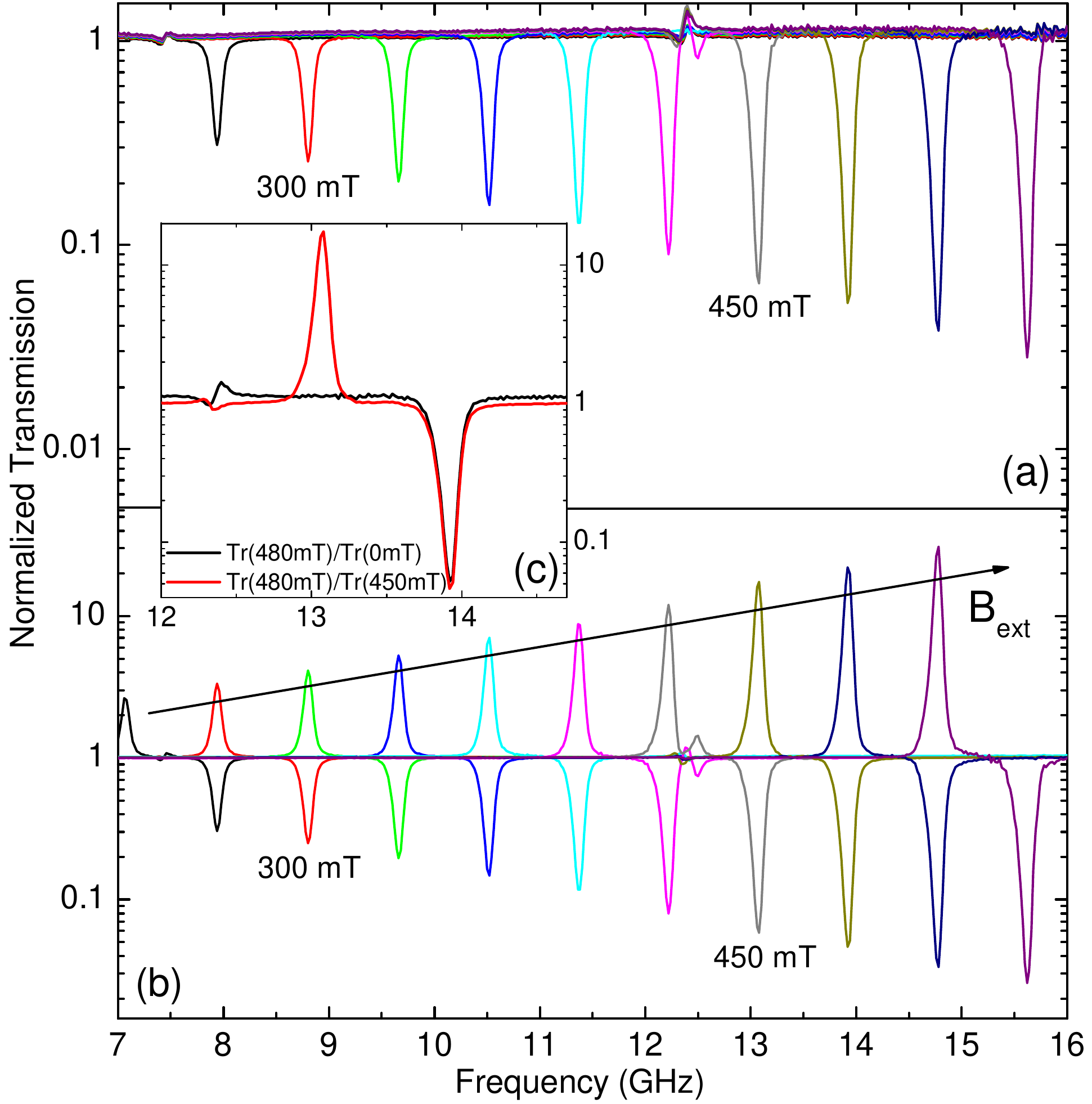}
	\caption{(color online) ESR spectra for NITPhOMe. (a) Broadband transmission spectra from $270$ to $540\,\milli\tesla$ in steps of $30\,\milli\tesla$ at $1.6\,\kelvin$ normalized with the zero-field spectrum. (b) Same raw data as in (a), but here the spectra are normalized to corresponding ones taken at 30\,mT smaller fields. (c) Comparison of the two different normalization methods at $480\,\milli\tesla$.}
	\label{fig:Figure2}
\end{figure}
Frequency-swept spectra of NITPhOMe are shown in Fig. \ref{fig:Figure2}(a) and (b) for two different normalizations. To identify the ESR signal from the raw data, all spectra are normalized with the zero-field spectrum (see Fig.\ \ref{fig:Figure2}(a)) or with a spectrum taken at a slightly different magnetic field (see Fig.\ \ref{fig:Figure2}(b)). One has to be careful that the positive peaks in the spectra are just artifacts of the normalization method and do not contribute to the actual ESR signal. Both techniques give clear spectra, and as evident from the direct comparison in Fig.\ \ref{fig:Figure2}(c), they result in ESR spectra which are basically identical. The increase of the ESR signal with increasing external magnetic field results from the stronger population difference of the Zeeman-split energy levels. In the previous study using superconducting coplanar lines \cite{Clauss13}, such a zero-field normalization was not viable because of the pronounced magnetic field dependence of microwave losses in superconductors, and only normalization like in Fig.\ \ref{fig:Figure2}(b) led to satisfactory results. Furthermore, the losses of Cu or Au at higher temperatures are substantially lower than those of good superconductors (such as Nb or Pb) \cite{Clauss13,Hafner14} when heated above their critical temperatures. Therefore, metallic coplanar lines are much more suitable than superconducting ones for broadband ESR studies of the type presented here.

\begin{figure}[tb]		
	\centering
	\includegraphics[width=8cm]{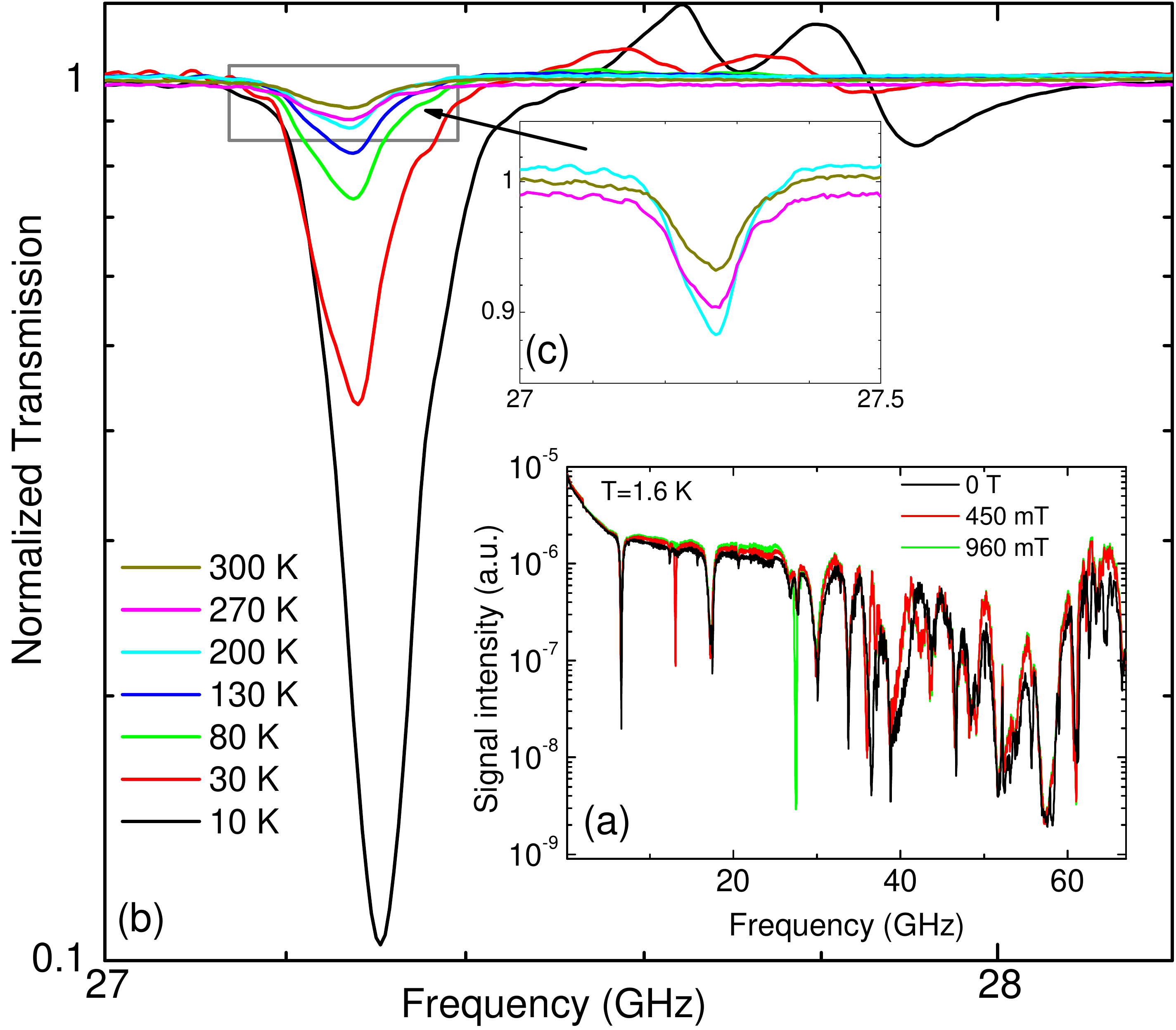}
	\caption{(color online) Temperature-dependent spectra of NITPhOMe for a magnetic field of $960\,\milli\tesla$ (a) Raw data: Broadband transmission spectra for three different magnetic fields (0, 450, 960 mT) at $1.6\,\kelvin$. (b) Normalized ESR signal for different temperatures. (c) Magnified view of the ESR for the highest measured temperatures.}
	\label{fig:Figure3}
\end{figure}
In Fig.\ \ref{fig:Figure3}(a), we illustrate raw spectra of the sample NITPhOMe for three different magnetic fields (0, 450 and $960\,\milli\tesla$) at $1.6\,\kelvin$. One can observe the typical decrease in transmission with increasing frequency due to frequency-dependent attenuation in the coaxial cables and the coplanar waveguide. The ESR is visible as strong transmission minima near 13~GHz (for 450~mT) and 27~GHz (for 960~mT). All other features arise from standing waves on the transmission line or in the box or are artifacts of the instruments. Fig.\ \ref{fig:Figure3}(b) shows temperature-dependent spectra of NITPhOMe. The ESR signal decreases with increasing temperature due to the thermal population difference but the ESR signal can be clearly identified up to $300\,\kelvin$ (Fig.\ \ref{fig:Figure3}(c)). A shift of the ESR absorption dip to higher frequencies is also observed for lower temperatures and this can be used to monitor the presence of intermolecular interactions. The sensitivity of our experiment at low temperatures is comparable to the former study using superconducting lines \cite{Clauss13}.
\begin{figure}[tb]		
	\centering
	\includegraphics[width=8cm]{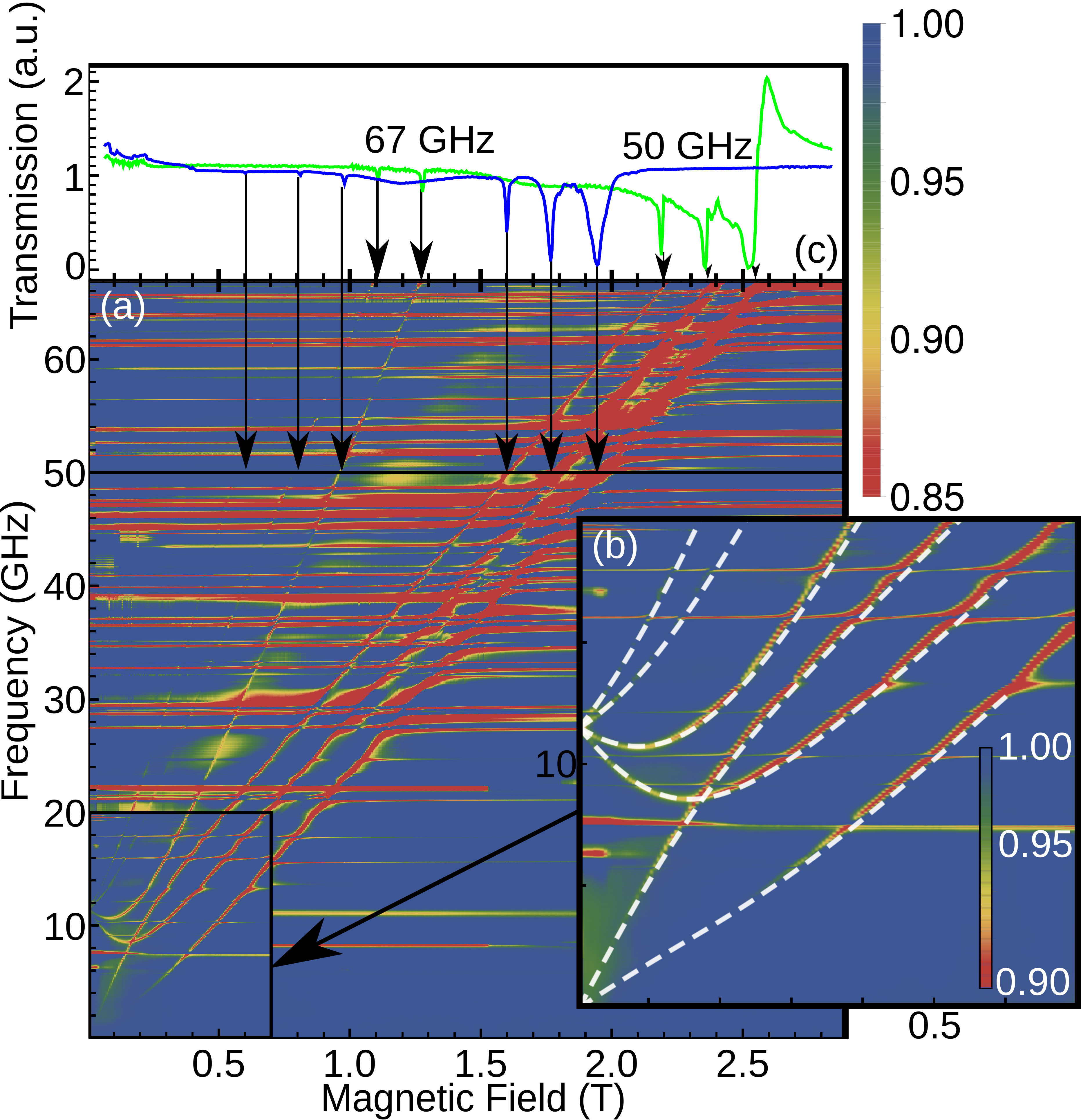}
	\caption{(color online) (a) Normalized frequency-swept ESR spectrum for ruby. ESR shows up as lighter areas. The transitions between the eigenfunctions of the Cr$^{3+}$ spin Hamiltonian can be seen. (b) Zoom of the transition spectrum from $0$ to $700\,\milli\tesla$ and a frequency range from $100\,\mega\hertz$ to $20\,\giga\hertz$. One can recognize the zero-field splitting. The white dashed lines indicate theoretical calculations. (c) Magnetic-field-swept spectra at $50$ and $67\,\giga\hertz$. The ESR signals of the six transitions are visible.}
	\label{fig:Figure4}
\end{figure}

We now move to the ruby sample which is a magnetic system with zero-field splitting and several ESR transitions \cite{Clauss13}. The frequency was swept between $100\,\mega\hertz$ and $67\,\giga\hertz$ whereas the magnetic field went up to $2.9\,\tesla$. For normalization, we used the average transmission of all magnetic fields for each particular frequency. Fig.\ \ref{fig:Figure4}(a) shows the characteristic six ESR transitions for ruby throughout a very broad frequency and field range. The horizontal lines in this plot are artifacts of the microwave measurements \cite{avoidedlevelcrossings}.
Fig.\ \ref{fig:Figure4}(b) shows in more detail the behavior for low fields and frequencies. One particularly recognizes the zero-field splitting around 11~GHz. The white dashed lines in Fig. \ref{fig:Figure4}(b) indicate theoretical calculations that fit well to our experimental data. Additionally, in Fig.\ \ref{fig:Figure4}(c) magnetic field-swept spectra are illustrated which show the ESR signals at very high frequencies of $50$ and $67\,\giga\hertz$.\\

\begin{figure}[tb]		
	\centering
	\includegraphics[width=8cm]{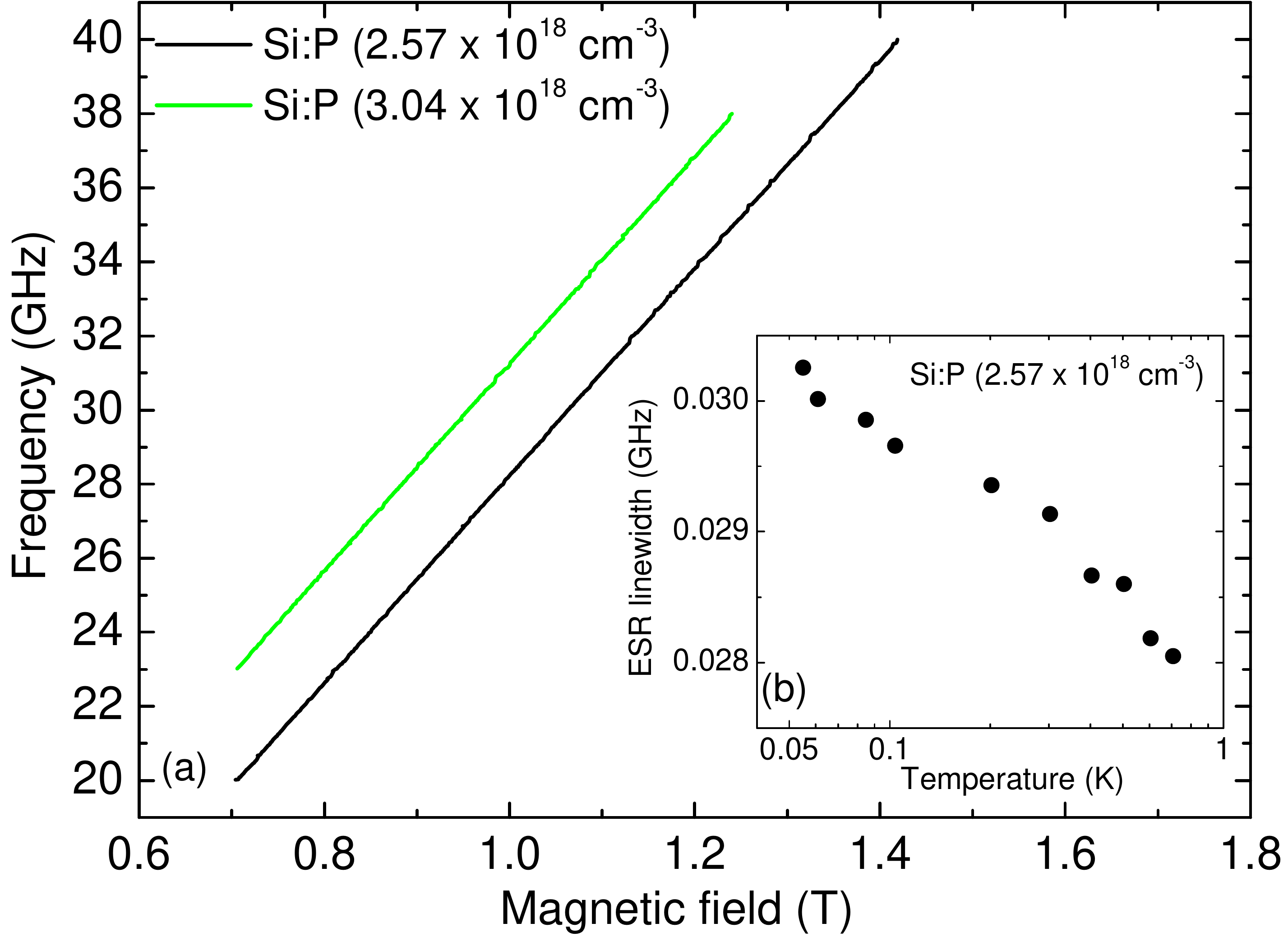}
	\caption{(color online) ESR measurements on doped Si:P. (a) ESR frequency as a function of the magnetic field at $1.6\,\kelvin$. The absorption dip minima were extracted from a normalized frequency-swept ESR spectrum like it is shown for ruby. For clarity, the data for the sample with higher doping has been shifted vertically by $3\,\giga\hertz$. (b) ESR linewidth in Si:P as a function of temperature at $640\,\milli\tesla$ performed in a dilution refrigerator.
	\label{fig:Figure5}}
\end{figure}
Our third sample material is Si:P, where the phosphorus dopants introduce spins that can be probed by ESR. Doped silicon is a prime candidate for spin-based quantum information processing in solids \cite{Wolfowicz13, Steger12, Pla12}, e.g.\ following the Kane proposal \cite{Kane1998, Morello10, Koenraad11}.
For our Si:P measurements we used a waveguide geometry with $S=100~\micro\meter$ and $W=41\,\micro\meter$.\\
Fig.\ \ref{fig:Figure5} plots the frequency of the absorption dip minimum as a function of the magnetic field for two different phosphorus concentrations ($2.57\times 10^{18}\,\centi\meter^{-3}$ and $3.04\times 10^{18}\,\centi\meter^{-3}$) \cite{Hering07} in a frequency range from $20$ to $40\,\giga\hertz$. Linear fits to those data reveal the $g$-value of the doped silicon  to be $g=1.99896\pm 0.00028$ ($2.57\times 10^{18}\,\centi\meter^{-3}$) and $g=1.99684\pm 0.00045$ ($3.04\times 10^{18}\,\centi\meter^{-3}$), respectively. The $g$-factor of Si:P has been studied intensely before, including the concentration dependence for heavily doped samples \cite{Feher1959,Quirt1973,Ochiai1976,Fasol01}. The inset of Fig.\ \ref{fig:Figure5} presents the ESR linewidth in $\giga\hertz$ as a function of temperature in Si:P doped to $2.57\times 10^{18}\,\centi\meter^{-3}$. Contrary to our other experiments, these measurements were performed in a $^3$He/$^4$He dilution refrigerator at sample temperatures down to 50~mK. One can recognize a rapid increase of the ESR linewidth with decreasing temperature. Former works on Si:P \cite{Quirt1973,Ue1971,Maekawa1965} have also reported the beginning of this trend for experiments down to $1\,\kelvin$ \cite{Quirt1973,Ue1971,Maekawa1965} and a more pronounced increase was found at mK temperatures \cite{Murayama1984,Mason1991}. This strong increase upon cooling was explained with exchange narrowing \cite{Murayama1984}. At this particular dopant density, one has a loss of the hyperfine structure of phosphorus due to an overlapping of the donor wavefunctions. So one obtains a motionally-narrowed line \cite{Abragam1961}. After our demonstration of broadband ESR in Si:P, one could now approach more complex related materials like Si:Bi \cite{Wolfowicz13}, but here sensitivity might be an issue.

In conclusion, we have shown a promising technique for ESR measurements in very broad frequency and magnetic field ranges by using metallic coplanar waveguides. During one single frequency sweep, we cover frequencies from $0.1$ to $67\,\giga\hertz$. This includes the conventional X and Q bands, but also extends to the highest frequencies that can still be accessed with coaxial cables for convenient broadband and cryogenic experiments. Depending on the particular goal, ESR can either be detected in frequency sweeps at a fixed magnetic field or in the more conventional field sweeps at fixed frequency. In the latter case, in contrast to commercial spectrometers, the frequency can be chosen at will, according to the scientific question. Furthermore, our compact coplanar ESR probe can easily be implemented in superconducting magnets and dilution refrigerators, thus opening up both magnetic fields and cryogenic temperatures that go well beyond other ESR techniques.

Accordingly, there is particularly large potential of this technique for ESR studies in those unconventional parameter spaces. Large magnetic fields in combination with comparably small ESR frequencies are relevant for field-induced level mixing with avoided crossings \cite{Gysler14}, and for such studies a very low sample temperature is also advantageous. Free choice of magnetic field and temperature is also relevant for ESR studies of materials that undergo magnetic phase transitions, here one example are quantum-critical heavy fermions with their characteristic energy scale in the mK range \cite{Scheffler13,Krellner08,Sichelschmidt03}, but also other magnetic materials with very low transition temperatures are of interest  \cite{Povarov11}. 
In all these cases, frequency-swept measurements can offer substantial advantages compared to field-sweeping as the measurement is performed at a fixed magnetic field, i.e.\ at a single point in the magnetic phase diagram.
We documented this wide applicability range by ESR measurements that span the cryogenic temperatures down to 50~mK for the Si:P example as well as temperatures up to room temperature for NITPhOMe, whereas ruby acts as an example with multiple transitions throughout a broad range of frequencies and magnetic field.

This work was supported by the Deutsche Forschungsgemeinschaft, including SFB/TRR 21 and SPP1601, the AvH Stiftung (Sofja Kovelevskaja award)  and the EU-FP6-COST Action MP1201. We thank G.\ Untereiner for experimental support, and M.\ Fanciulli for stimulating discussions.


\begin{thebibliography}
\bibitem{filmorientation}The design of the experimental insert allows a film orientation parallel to the magnetic field. We cannot exclude a misalignment of the field but we assume a tilting out of plane of less than 2\degree.
\bibitem{power}Due to frequency-dependent attenuation in the coaxial cables and in the coplanar line, the transmission decreases with increasing frequency. Therefore, the input power was swept linearly from $-20~\deci\bel\milli$ to $10~\deci\bel\milli$ for NITPhOMe and ruby. The temperature-dependent measurements on NITPhOMe were performed at a constant input power of $P_{\rm in}=-10 \deci\bel\milli$. For the doped Si:P, a constant value of $-20~\deci\bel\milli$ was used.
\bibitem{avoidedlevelcrossings}These horizontal lines stem from undesired resonant modes within the sample box or the overall microwave assembly. The particular, thick ruby sample fills a substantial fraction of the sample box volume and thus can interact with such parasitic microwave box modes, leading to anticrossings in Fig. \ref{fig:Figure4}.
\end{thebibliography}
\end{document}